\documentclass[%
 reprint,
superscriptaddress,
 amsmath,amssymb,
 aps,
]{revtex4-1}

\usepackage[pdftex]{graphicx} 
\usepackage{dcolumn}
\usepackage{bm}

\begin{document}

\preprint{APS/123-QED}

\widetext
\leftline{Primary authors: T. Okamoto}
\leftline{Comment to {\tt okamoto.t.ap@m.titech.ac.jp}}


\title{Topological effects of three-dimensional porous graphene on Dirac quasiparticles}

\author{Takuya Okamoto}
\email{okamoto.t.ap@m.titech.ac.jp} 
\affiliation{Department of Electrical and Electronic Engineering, Tokyo Institute of Technology, 2-12-1 Ookayama, Meguro-ku, Tokyo 152-8552, Japan}
\affiliation{Laboratory for Future Interdisciplinary Research of Science and Technology, Tokyo Institute of Technology, 2-12-1, Ookayama, Meguro-ku, Tokyo 152-8552, Japan}
\author{Yoshikazu Ito}
\affiliation{Institute of Applied Physics, Graduate School of Pure and Applied Sciences, University of Tsukuba, Tsukuba, Ibaraki 305-8573, Japan}
\author{Naoka Nagamura}
\affiliation{National Institute for Materials Science, 1-2-1 Sengen, Tsukuba, Ibaraki 305-0047, Japan}
\affiliation{Japan Science and Technology Agency, PRESTO, 4-1-8, Honcho, Kawaguchi, Saitama, 332-0012, Japan}
\author{Keishi Akada}
\affiliation{Institute for Solid State Physics, The University of Tokyo, Kashiwa, Chiba, 277-8581, Japan}
\author{Takeshi Fujita}
\affiliation{School of Environmental Science and Engineering, Kochi University of Technology, 185 Miyanokuchi, Tosayamada, Kami City, Kochi 782-8502, Japan}

\author{Yukio Kawano}
\email{kawano@pe.titech.ac.jp}
\altaffiliation[permanent address: ]{Laboratory for Future Interdisciplinary Research of Science and Technology, Tokyo Institute of Technology, 2-12-1, Ookayama, Meguro-ku, Tokyo 152-8552, Japan}
\affiliation{Department of Electrical and Electronic Engineering, Tokyo Institute of Technology, 2-12-1 Ookayama, Meguro-ku, Tokyo 152-8552, Japan}
\affiliation{Laboratory for Future Interdisciplinary Research of Science and Technology, Tokyo Institute of Technology, 2-12-1, Ookayama, Meguro-ku, Tokyo 152-8552, Japan}


  \begin{abstract}
   This paper reports on the topological effects of three-dimensional (3D) porous graphene with tunable pore sizes and a preserved 2D graphene system of Dirac quasiparticles on its electrical properties.
   This 3D architecture is characterized by the intrinsic curvature of smoothly interconnected graphene sheets without edges, the structures and properties of which can be controlled with its pore sizes.
   The impact of pore size on the electrical transport properties was investigated through magnetoresistance measurements.
   We observed that 3D graphene with small pores exhibits transitioning to weak localization with decreasing temperature.
   The comparison with the theory based on the quantum correction clarified that an increase in the intrinsic curvature significantly induces the intervalley scattering event, which breaks the chirality.
   This increase in the intervalley scattering rate originates from the unique topological effects of 3D graphene, i.e., the topological defects required to form the high curvature and the resulting chirality mixing.
   We also discuss the scattering processes due to microscopic chemical bonding states as found by high spatial-resolved X-ray photoemission spectral imaging, to support the validity of our finding.
  \end{abstract}

\maketitle 

\section{Introduction} 
Graphene, with a two-dimensional (2D) hexagonal-carbon lattice, has unique electrical properties characterized by Dirac quasiparticles with a chiral nature, and thus provides many intriguing applications in the field of physics \cite{1,2,3}.
An interesting property is the insensitivity of the chirality to long-range scattering potential on a length scale larger than the carbon-lattice spacing, providing strong independence of the valley state \cite{2}.
In addition, as graphene is resistant to perpendicular strain, its structure can spread to the three-dimensional (3D) space with the introduction of a negative Gaussian curvature \cite{6,7,8,9,10}.
This 3D curved graphene is a very interesting system, wherein Dirac quasiparticles captured in the 2D curved space behave under 3D topology.
The first prediction of this system was theoretically proposed in \cite{6}, and was referred to as “3D graphene.”
One of the remarkable features of 3D graphene is the presence of intrinsic curvature over large length scales to maintain the spatial shape of the graphene sheet in the 3D space, while preserving the chirality of the Dirac quasiparticles. Another important aspect is the appearance of non-hexagonal carbon rings (i.e., topological defects), such as heptagonal ring and pentagon–heptagon pair (dislocation), which are inherently included in the hexagonal lattice to allow stable deviation from 2D planarity according to Euler$'$s theorem \cite{7,8,9,10,11}.
As high amounts of such topological defects are required to form a highly curved sheet, the macroscopic topology can be engineered through the proper atomic design of the intrinsic curvature.
Interestingly, it is known that even though the range of its characteristic defect potential is larger than the lattice spacing, topological defects significantly impact the chirality of Dirac quasiparticles through a topological effect \cite{2}.
These aforementioned features are expected to provide rich curvature-dependent physical properties, which are important for the manifestation of topological effects on 3D graphene.
Owing to the strong potential of this emerging material, this system has been primarily studied theoretically \cite{7,8,9}.

       As quantum interference, such as weak localization (WL), strongly depends on a phase of electronic wavefunction, it can be employed as a useful probing technique to derive clear information on the electronic states and scattering processes \cite{12}.
        For conventional 2D systems, only inelastic scattering processes are known to be relevant to WL \cite{12}, whereas both inelastic and elastic scattering processes are responsible for the chirality in graphene \cite{13,14}.
	 Because the electronic structures of 3D graphene can be controlled by engineering the intrinsic curvature, its unique electrical properties, such as scattering type and rate, should be experimentally explored using WL measurements.
	  
	   In this paper, we experimentally present the intrinsic-curvature effects on the electrical properties of 3D graphene through WL measurements.
	    As a 3D graphene system, we utilized high-quality 3D porous graphene \cite{15}, which has an open porous structure characterizing its intrinsic curvature based on a smoothly interconnected graphene network.
	     We observed that an increase in the intrinsic curvature induces transitioning to a localized electrical system through geometric comparison. Our study demonstrates that the design of a 3D topology enables the control of the properties of Dirac quasiparticles.

\section{Result and Discussion}
\subsection{Material characterizations}
\begin{figure}[tbp]
 \centering
  \includegraphics[width=90mm]{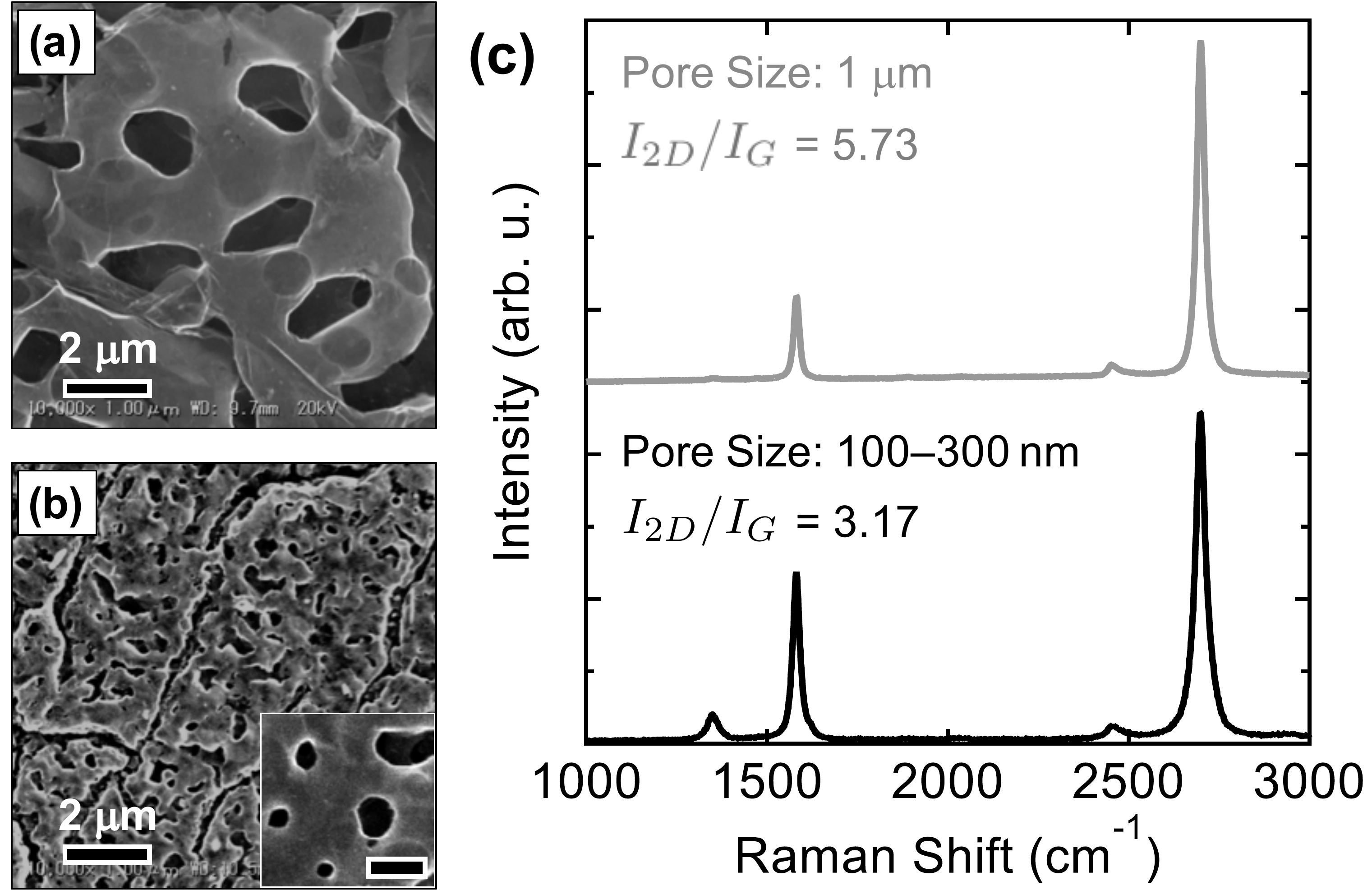}
 \caption{ Scanning electron microscopy images of 3D porous graphene with (a) large pores and (b) small pores.
 Inset to (b) shows the enclosed image with a scale bar of 200 nm. (c) Raman spectra of samples.
 (d) Optical image of sample used in electrical transport measurements.}
    \label{fig1}
\end{figure}
The 3D porous graphene samples were synthesized on a porous Ni template by using a chemical vapor deposition (CVD) method, in which the intrinsic curvature of the porous structure was tuned by controlling the CVD conditions \cite{15}.
After dissolving the Ni template and coon the pore size; the peak intensity of the small-pore-sized sample is 10 times greater than that of the large-pore-sized sample (Table SI).
This strong pore dependence suggests the existence of topological defects of the geometrically dependent 3D architecture \cite{15,17,18,19}.
The detailed analysis of the topological defects, including the pore-size dependence of their amount, is shown in Fig. \ref{fig4}, which presents the chemical origin.

 \subsection{Electrical transport measurements}
 \begin{figure}[tbp]
 \centering
  \includegraphics[width=90mm]{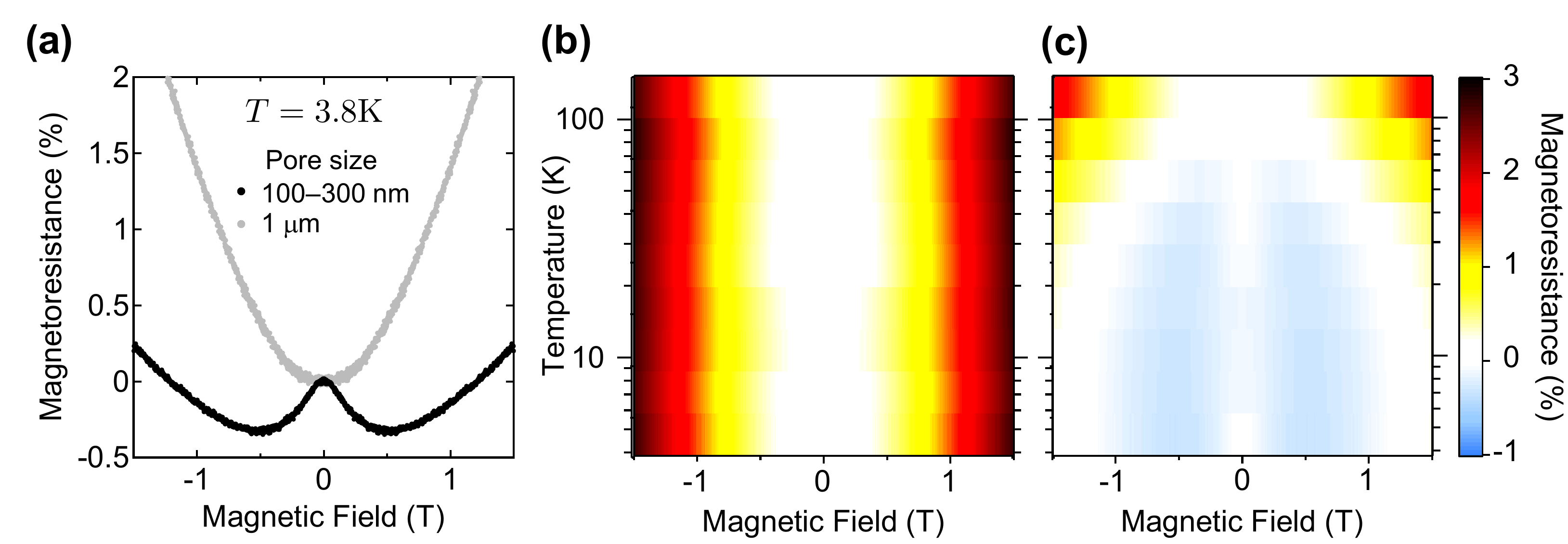}
  \caption{ (a) Pore-sized dependence of normalized MR at 3.8 K.
  (b) Color plot of normalized MR of large-pore-sized sample as a function of $B$ and $T$;
  (c) small-pore-sized sample. The temperature dependence was smoothed.}
      \label{fig2}
\end{figure}
The electrical transport properties of the 3D porous graphene were then studied using a standard four-probe method to avoid the influence of contact resistance.
Figure \ref{fig2}(a) plots the normalized longitudinal MR of $\Delta R_{xx}(B)/R_{xx}(0)$, where $\Delta R_{xx}(B) = R_{xx}(B) -  R_{xx}(0)$, for the two samples under an external magnetic field range of $|B| \leq 1.5$ T at temperature $T = 3.8$ K.
As the observed MR shows asymmetric behaviors for the polarity of $B$ (Fig. S2), we employed  $R_{xx}(B) = \left\{R_{xx}(B) + R_{xx}(-B)\right\}/2$ (see supplemental materials).
The large-pore-sized sample was observed to exhibit positive MR ($\Delta R_{xx} > 0$), whereas the small-pore-sized sample showed negative MR ($\Delta R_{xx} < 0$).
These results indicate that the intrinsic curvature, which is determined by the pore size, strongly affects the electrical transport properties of the 3D porous graphene.
We further measured the temperature ($T$) dependence of the MR and plotted the normalized MR as a function of $B$ and $T$ in Figs. \ref{fig2}(b) and (c).
The large-pore-sized sample exhibits almost no $T$ dependence, indicating that quantum correction on the MR is irrelevant and classical transport is dominant over all the measured $T$ ranges.
Indeed, our data are in good agreement with the classical transport model of cyclotron motion \cite{26}, shown in Fig. S3.
In contrast, the small-pore-sized sample showed strong $T$ dependence, indicating that the negative MR arises from WL induced by the quantum correction on the electronic phase.
This quantum correction gradually disappears with an increase in the temperature of up to a few tens of kelvins; this is a typical behavior of WL in graphene \cite{20,21,22,23}.

\begin{figure}[tbp]
 \centering
  \includegraphics[width=90mm]{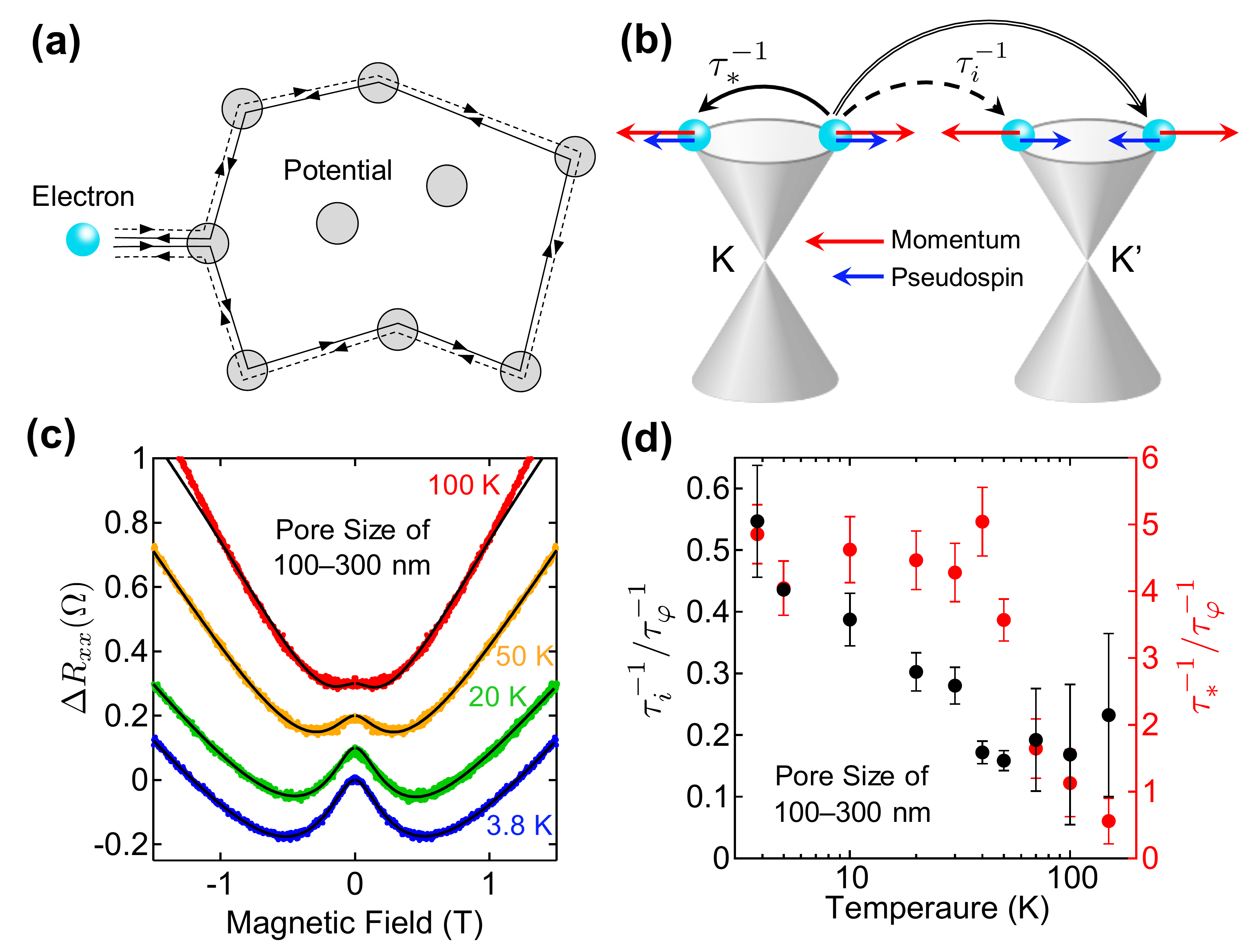}
 \caption{(a) Trajectories of an electron scattered by potentials, which contribute to WL correction.
 (b) Schematic of band structure of graphene. The dotted and solid arrows represent intra- and inter-valley scattering processes, respectively.
 The double-line arrow represents chirality mixing induced by topological defects.
 (c) $\Delta R_{xx}$ of the small-pore-sized sample at typical temperatures of 3.8 K (blue dots), 20 K (green dots), 50 K (orange dots), and 100 K (red dots).
 Each curve is shifted upward for clarity.
 The lines are best fits to Eq. \ref{eq}.
 (d) Temperature dependence of the characteristic ratio of scattering rates $\tau_{i}^{-1}/\tau_{\varphi}^{-1}$ and $\tau_{*}^{-1}/\tau_{\varphi}^{-1}$.}
      \label{fig3}
\end{figure}

	Next, we discuss the WL observed for the small-pore-sized sample.
	The WL is based on the quantum interference of electron waves.
	Figure \ref{fig3}(a) illustrates the trajectories of the electrons scattered by potential, leading to the WL correction on the resistance.
	As the two electrons follow the same pathway, the phase of their electron waves is identical to that in conventional 2D systems, such as silicon metal-oxide-semiconductor field-effect transistors and gallium arsenide high-electron-mobility transistors; accordingly, they positively interfere with each other.
	As this interference is broken by applying a magnetic field, the effect of the WL correction manifests itself as the occurrence of negative MR in the low magnetic field \cite{12}.
	For the graphene system, the above-mentioned process differs somewhat from that of a conventional 2D system owing to an additional internal degree of freedom \cite{13}, known as valley index (K and K'{} points) and pseudospin [Fig. \ref{fig3}(b)].
	This directly implies that the WL correction on the graphene system depends not only on the inelastic scattering rate ($\tau_{\varphi}^{-1}$) but also on the elastic scattering rate related to the valley index and pseudospin, which consist of inter- and intra-valley scattering rates ($\tau_{i}^{-1}$, $\tau_{*}^{-1}$), respectively.
	These scattering processes are represented by the solid and dotted arrows in Fig. \ref{fig3}(b).
	Based on the above-mentioned discussion, the resistivity correction, $\Delta \rho_{xx}(B)$, can be expressed as \cite{13}
	\begin{align}
	 \varDelta \rho_{xx}(B) &= - \frac{e^2 \rho_{xx}^{2}}{\pi h} \left\{F\left(\frac{\tau_{B}^{-1}}{\tau_{\varphi}^{-1}}\right) \right. \notag \\ 
	  &	    \left. - F\left(\frac{\tau_{B}^{-1}}{\tau_{\varphi}^{-1}+2\tau_{i}^{-1}}\right) - 2F\left(\frac{\tau_{B}^{-1}}{\tau_{\varphi}^{-1}+\tau_{*}^{-1}}\right)\right\}
	   \label{eq}
	   \end{align}
	   where $F(x) = \ln{x} + \Psi (0.5 + x^{-1})$ and $\tau_{B, \varphi , i, *}^{-1}=4eDB_{\varphi , i, *}/ \hbar$, $\Psi (x)$ is the digamma function, $e$ is the elementary charge, $D$ is the diffusion constant, and $\hbar$ is the reduced Planck$'$s constant.
	   Equation (\ref{eq}) simply indicates that a large $\tau_i^{-1}$ value leads to negative MR (WL) but is oppositely compensated by an increase in $\tau_*^{-1}$ or in $\tau_{\varphi}^{-1}$.
	   The application of this equation as the fitting curve to the experimental data enables the derivation of these scattering rates. 
	   
	   We fitted the MR data using Eq. (\ref{eq}) for the small-pore-sized sample in the magnetic-field range of $|B| \leq  1.5$ T, where the WL prevails.
	   As it is difficult to accurately determine the channel length, width, and thickness of the porous graphene, we introduced a dimensionless fitting parameter, $\alpha \sim$ 31.2, as $\rho_{xx} = \alpha R_{xx}$.
	   The experimental results agree well with the fitted curves [black solid lines in Fig. \ref{fig3}(b)], deriving the characteristic scattering rates responsible for the WL: $\tau_i^{-1}/\tau_{\varphi}^{-1}$ and $\tau_*^{-1}/\tau_{\varphi}^{-1}$.
	   These ratios are plotted as a function of temperature $T$ in Fig. \ref{fig3}(d).
	   For the low temperature region, $\tau_{i}^{-1}$ and $\tau_{*}^{-1}$ contribute significantly to the scattering processes; this reveals that despite a large $\tau_{*}^{-1}$ value, the WL occurs because of the significant inter-valley scattering, i.e., $\tau_{i}^{-1} \sim \tau_{\varphi}^{-1}$.

	   First, the most important and interesting result is the significant inter-valley scattering rate.
	   In general, this scattering process requires large changes in momentum, as depicted in Fig. \ref{fig3}(b), and hence is induced by strong short-range potentials shorter than the lattice space, such as edge, tight coupling with the substrate, and point defects \cite{12,13,21,22,23}.
	   However, as the 3D porous graphene is a free-standing structure with a vast surface area, the contributions of the substrate and edge are negligible, as depicted in Figs. \ref{fig1}(a), (b), and (d).
	   In addition, the amounts of point defects are trivial because the high temperature (800 -- 900 $^\circ$C) of the CVD process reconstructs such defects as dislocations.
	   Therefore, the short-range potential is not a dominant mechanism affecting inter-valley scattering events.
	   By considering that the inter-valley scattering promotes chirality mixing, as shown in Fig. \ref{fig3}(b), another chirality-mixing mechanism should be included. Geometric comparisons of the D-band in the Raman spectra and MR behaviors show that the topological defects are anticipated to be a critical factor.
	   The chirality mixing caused by the topological defects has been reported for other graphene systems as a topological effect, e.g., fullerene \cite{27} and graphene cones \cite{17,18,24}.
	   When electrons traverse a path that encloses topological defects, both pseudospin and valley index are interchanged \cite{2,17,18,24}.
	   Based on these physical findings, we interpret that low-temperature transport properties of our 3D porous network graphene are characterized by the chirality mixing, indicated by the double-line arrow in Fig. \ref{fig3}(b) and represented by the sum of the inter-valley and intra-valley scattering, because of the absence of large momentum changes.
	   Another important point is that this topological effect does not originate from short-range potential associated with large momentum conversion, but from pure topological effects induced by the presence of the topological defects, which is a characteristic of the 3D graphene with a deviation from flat 2D geometry.
	   These results offer an interesting possibility that the electrical properties of 3D graphene, especially chirality, can be manipulated by designing and engineering the graphene’s intrinsic curvature.
	   
	   Second, the intra-valley scattering was observed to be predominant for the small-pore-sized sample at low temperatures, as depicted in Fig. \ref{fig3}(d).
	   As the above-mentioned topological effect is approximately equivalent to $\tau_{i}^{-1} + \tau_{*}^{-1}$, there should exist several other possible intra-valley scattering events \cite{24}.
	   This scattering process with small changes in momentum originates from long-range potential, such as ripple, line defects (e.g., dislocation and grain boundary), distortions induced by topological defects, and electrostatic potentials induced by charged impurities \cite{2, 12, 13, 21,24}.
	   Although the precise determination of the origin of the scattering process is difficult, the long-range distortion is one of the possible mechanisms due to the characteristic intrinsic curvature of the 3D graphene.
	   To experimentally confirm this speculation, we attempted to obtain information on the distortion through chemical-state-selective observations using scanning photoelectron microscopy, which can obtain the spatial distribution of the XPS information \cite{28}.
	   In the case of 3D graphene, the sp$^2$ bonding character diminishes and the sp$^3$ bonding character increases (sp$^2$/sp$^3$ hybridization) in the curved region of the 3D architecture \cite{25,26,27}.
	   By using this high-spatial-resolution XPS, we performed microscopic studies on the contribution of intrinsic curvature to the electrical properties, instead of using the average information acquired from the previous electrical transport measurements.

\subsection{Nanoscale X-ray photoemission spectroscopy.}
\begin{figure}[tbp]
 \centering
 \includegraphics[width=80mm]{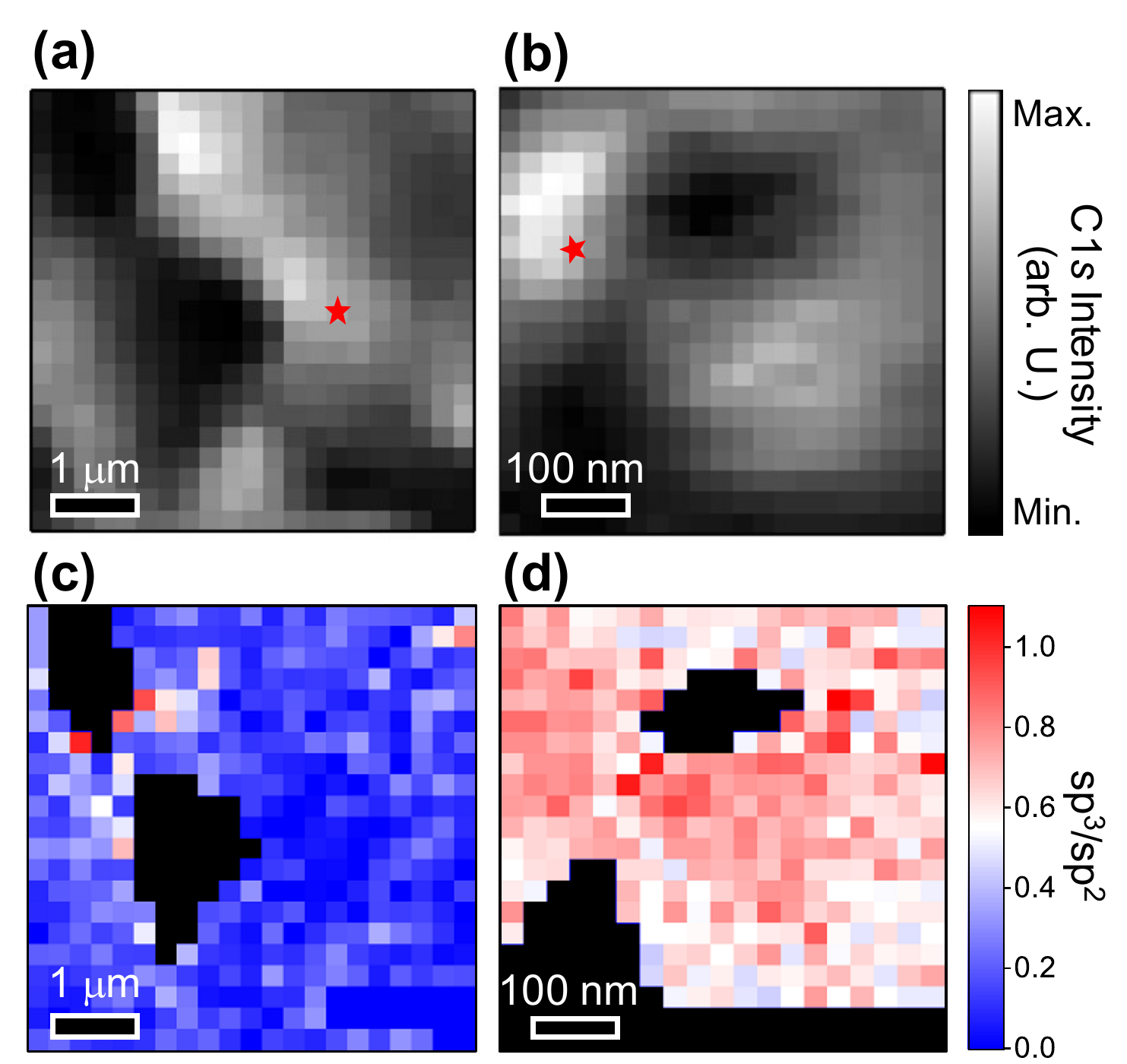}
 \caption{Mapping images of results obtained through high spatial-resolution XPS with 21 $\times$ 21 pixels.
 (a) C 1$s$ core-level intensity of large-pore-sized sample;
 (b) small-pore-sized sample.
 The C 1$s$ spectra, obtained at the point marked by red stars, are plotted in Fig. S4.
 Ratio sp$^3$/sp$^2$ of (c) large-pore-sized sample and (d) small-pore-sized sample.
 The areas of pores are crossed out due to the background signal.}
      \label{fig4}
\end{figure}
Figures \ref{fig4}(a) and (b) map out the distribution image of C 1$s$ intensity for the large- and small-pore-sized samples corresponding to each topography.
These C 1$s$ spectra can be divided into two components through fitting analysis (Fig. S4): dominant intensity sp$^2$ \cite{29} and smaller intensity sp$^3$ \cite{30}.
The spatial distribution of the bonding-character ratio, sp$^3$/sp$^2$, is illustrated in Figs. \ref{fig4}(c) and (d).
The small-pore-sized sample is clearly sp$^3$-richer than the large-pore-sized sample.
This finding also demonstrates that 3D topology modifies the electrical properties through the chemical nature.
For the large-pore-sized sample, the chemical bonding nature is locally modified to be sp$^3$-rich in the high-curvature regions around the pores, as clearly depicted in the overlay image in Fig. S5.
However, for the small-pore-sized sample, the chemical bonding nature does not correlate with the pores, i.e., it is strongly modified to be sp$^3$-rich in the entire observed region (Fig. S5).
This result indicates that the distortion formed by topological defects may be spread throughout the sample.
Such a distortion with topological defects induces intra- and inter-valley scattering events, as mentioned in the above-mentioned MR analysis.
These findings are consistent with the electrical transport properties and suggest that the distortion is one of the main mechanisms of the intra-valley scattering process \cite{3,13,21,22}. 
\section{CONCLUSION}
In summary, we investigated the 3D topological effects of the intrinsic curvature of 3D graphene networks with a tunable porous structure on its electrical properties.
The MR measurements demonstrated that the induction of the high curvature transformed the 3D porous graphene states from classical to WL states.
The geometrical comparison revealed that this occurrence is due to topological effects originating from topological defects inherently required for the formation of 3D architecture with deviation from the flat 2D geometry.
The additional XPS measurements confirmed that the higher curvature modifies the sp$^3$-rich bonding nature, thus demonstrating the existence of distortion and topological defects due to the 3D architecture with deviation from the flat 2D geometry.
This work opens up new possibilities in the area of topological physics of 3D structures composed of the family of the other 2D materials.




 \section{acknowledgments}
 We are grateful to Prof. M. Koshino of Osaka University for theoretical discussions on the experimental data, and Prof. M. Hatano of Tokyo Institute of Technology for permission to use Raman spectroscopy.
 This work was supported in part by the JST-Mirai Program, Matching Planner Program, Center of Innovation Program, and “Materials Research by Information Integration” Initiative project of PRESTO (Grant Number: JPMJPR17NB) from the Japan Science and Technology Agency; JSPS KAKENHI Grant Numbers JP16J09937, JP17K19026, JP17H02730, JP16H00798, JP16H00906, JP18H03766, JP18H04477, 18K14174 and 19J15245 from the Japan Society for the Promotion of Science, the Murata Science Foundation, and Support for Tokyo Tech Advanced Researchers (STAR), JSPS Grant-in-Aid for Scientific Research on Innovative Areas “Discrete Geometric Analysis for Materials Design”; Grant Number JP18H04477.
 This work was performed at the Synchrotron Radiation Research Organization, University of Tokyo (SPring-8 Proposal Nos. 7580 for 2018 and 7451 for 2019).


\begin{thebibliography}{99}
\bibitem{1}K.S. Novoselov, A.K. Geim, S.V. Morozov, D. Jiang, Y. Zhang, S.V. Dubonos, I.V. Grigorieva, and A.A. Firsov, “Electric field effect in atomically thin carbon films.” Science 306 (5696) (2004): 666–669.
 \bibitem{2}A. H. Castro Neto, F. Guinea, N. M. R. Peres, K. S. Novoselov, and A. K. Gei,"The electronic properties of graphene." Reviews of modern physics 81(1) (2009): 109.
 \bibitem{3}A.K. Geim and K.S. Novoselov, “The rise of graphene.” Nanoscience and Technology: A Collection of Reviews from Nature Journals. 2010. 11–19.
 \bibitem{4}H.W. Kroto, J.R. Heath, SC O'Brien, RF Curl, RE Smalley. “C60: Buckminsterfullerene.” Nature 318(6042) (1985): 162.
 \bibitem{5}S. Iijima and T. Ichihashi. “Single-shell carbon nanotubes of 1-nm diameter.” Nature 363(6430) (1993): 603.
\bibitem{6}A.L. Mackay and H. Terrones. “Diamond from graphite.” Nature 352(6338) (1991): 762.
\bibitem{7}H. Aoki, M. Koshino, D. Takeda, H. Morise, and K. Kuroki, “Electronic structure of periodic curved surfaces—continuous surface versus graphitic sponge.” Physica E: Low-dimensional Systems and Nanostructures 22(1–3) (2004): 696–699.
\bibitem{8}H. Weng, Y. Liang, Q. Xu, R. Yu, Z. Fang, X. Dai, and Y. Kawazoe, “Topological node-line semimetal in three-dimensional graphene networks.” Physical Review B 92(4) (2015): 045108.
\bibitem{9}M. Koshino and H. Aoki, “Dirac electrons on three-dimensional graphitic zeolites: A scalable mass gap.” Physical Review B 93(4) (2016): 041412.
\bibitem{10}M. Tagami, Y. Liang, H. Naito, Y. Kawazoe, and M. Kotani. “Negatively curved cubic carbon crystals with octahedral symmetry.” Carbon 76 (2014): 266–274.
\bibitem{11}H. Kusumaatmaja and D.J. Wales. “Defect motifs for constant mean curvature surfaces.” Physical Review Letters 110(16) (2013): 165502.
\bibitem{12}S. Hikami, A.I. Larkin, and Y. Nagaoka. Spin-orbit interaction and magnetoresistance in the two dimensional random system: Progress of Theoretical Physics 63 (1980): 707–710.
\bibitem{13}E. McCann, K. Kechedzhi, V.I. Fal’ko, H. Suzuura, T. Ando, and B. Altshuler, “Weak-localization magnetoresistance and valley symmetry in graphene.” Physical Review Letters 97(14) (2006): 146805.
\bibitem{14}V. I. Fal’ko, K. Kechedzhi, E. McCann, B. Altshuler, H. Suzuura, and T. Ando,” Weak localization in graphene,” Solid State Communications 143(33) (2007).
\bibitem{15}Y. Ito, Y. Tanabe, H-J Qiu, K. Sugawara, S. Heguri, N.H. Tu, K.K. Huynh, T. Fujita, T. Takahashi, K. Tanigaki, and M. Chen, “High‐quality three‐dimensional nanoporous graphene,” Angewandte Chemie International Edition 53(19) (2014): 4822–4826. 
\bibitem{16}A.C. Ferrari, J.C. Meyer, V. Scardaci, C. Casiraghi, M. Lazzeri, F. Mauri, S. Piscanec, D. Jiang, K. S. Novoselov, S. Roth, and A. K. Geim, “Raman spectrum of graphene and graphene layers.” Physical Review Letters 97(18) (2006): 187401.
\bibitem{17}K. Sasaki, Y. Sekine, K. Tateno,  and H. Gotoh, “Topological Raman band in the carbon nanohorn.” Physical Review Letters 111(11) (2013): 116801.
\bibitem{18}K. Sasaki, Y. Tokura, and T. Sogawa. “The origin of Raman D band: bonding and antibonding orbitals in graphene.” Crystals 3(1) (2013): 120–140.
\bibitem{19}Y. Ito, Y. Shen, D Hojo, Y. Itagaki, T. Fujita, L. Chen, T Aida, Z Tang, T. Adschiri, and M. Chen, “Correlation between chemical dopants and topological defects in catalytically active nanoporous graphene.” Advanced Materials 28(48) (2016): 10644–10651.
\bibitem{20}S. V. Morozov, K. S. Novoselov, M. I. Katsnelson, F. Schedin, L. A. Ponomarenko, D. Jiang, and A. K. Geim, “Strong suppression of weak localization in graphene.” Physical Review Letters 97(1) (2006): 016801.
\bibitem{21}F. V. Tikhonenko, D. W. Horsell, R. V. Gorbachev, and A. K. Savchenko, “Weak localization in graphene flakes.” Physical Review Letters 100(5) (2008): 056802.
\bibitem{22}F. V. Tikhonenko, A. A. Kozikov, A. K. Savchenko, and R. V. Gorbachev, “Transition between electron localization and antilocalization in graphene.” Physical Review Letters 103(22) (2009): 226801.
\bibitem{23}H. Cao, Q. Yu, L.A. Jauregui, J. Tian, W. Wu, Z. Liu, R. Jalilian, D. K. Benjamin, Z. Jiang, J. Bao, S. S. Pei, and Yong P. Chen, “Electronic transport in chemical vapor deposited graphene synthesized on Cu: Quantum Hall effect and weak localization.” Applied Physics Letters 96(12) (2010): 122106.
\bibitem{24}A. F. Morpurgo and F. Guinea. “Intervalley scattering, long-range disorder, and effective time-reversal symmetry breaking in graphene.” Physical Review Letters 97(19) (2006): 196804.
\bibitem{25}X. Wu, X. Li, Z. Song, C. Berger, and W. A. de Heer. “Weak antilocalization in epitaxial graphene: Evidence for chiral electrons.” Physical Review Letters 98(13) (2007): 136801.
\bibitem{26}Y. Tanabe, Y. Ito, K. Sugawara, D. Hojo, M. Koshino, T. Fujita, T. Aida, X. Xu, K. K. Huynh, H. Shimotani,  T. Adschiri, T Takahashi, K. Tanigaki, H. Aoki, and M Chen, “Electric properties of Dirac fermions captured into 3D nanoporous graphene networks.” Advanced Materials 28(46) (2016): 10304–10310.
\bibitem{27}J. González, F. Guinea and M.A. Vozmediano. “Continuum approximation to fullerene molecules.” Physical Review Letters. 69(1) (1992): 172.
\bibitem{28}K. Horiba, Y. Nakamura, N. Nagamura, S. Toyoda, H. Kumigashira, M. Oshima, K. Amemiya, Y. Senba, and H. Ohashi, “Scanning photoelectron microscope for nanoscale three-dimensional spatial-resolved electron spectroscopy for chemical analysis” Review of Scientific Instruments 82(11) (2011): 113701.
\bibitem{29}A. Pirkle, J. Chan, A. Venugopal, D. Hinojos, C. Magnuson, S. McDonnell, L. Colombo, E. Vogel, R. Ruoff, and R. Wallace, “The effect of chemical residues on the physical and electrical properties of chemical vapor deposited graphene transferred to SiO2.” Applied Physics Letters 99 (2011): 122108.
\bibitem{30}J. Diaz, G. Paolicelli, S. Ferrer, and F. Comin, “Separation of the sp$^3$ and sp$^2$ components in the C1$s$ photoemission spectra of amorphous carbon films” Physical Review B 54 (1996): 8064.
\end{thebibliography}

\end{document}